\documentstyle[12pt]{article}
\begin{document}

\title{Reaction rate for two--neutron capture\\ by $^4$He}

\author{V.D. Efros\thanks{Permanent address: Kurchatov Institute, 
Institute for General and 
Nuclear Physics, SU--123182 Moscow, Russia},
W. Balogh, H. Herndl, R. Hofinger\\ and H. Oberhummer\\
Institute of Nuclear Physics,\\ Technical University Vienna,\\
Wiedner Hauptstr. 8--10, A--1040 Wien, Austria}

\date{ }
\maketitle

\begin{abstract}
Recent investigations suggest that the
neutrino--heated hot bubble between the nascent neutron star and
the overlying stellar mantle of a type--II supernova may
be the site of the r--process. In the preceding $\alpha$--process
building up the elements to $A \approx 100$, the 
$^4$He(2n,$\gamma$)$^6$He-- and $^6$He($\alpha$,n)$^9$Be--reactions bridging
the instability gap at $A=5$ and $A=8$ could be of relevance. We suggest a
mechanism for $^4$He(2n,$\gamma$)$^6$He and calculate the
reaction rate within the $\alpha$+n+n approach. The value
obtained is about a factor 1.6 smaller than the one obtained
recently in the simpler direct--capture model, but is at least three order
of magnitude enhanced compared to the previously adopted value. Our calculation
confirms the result of the direct--capture calculation that
under representative conditions in the $\alpha$--process the reaction path
proceeding through $^6$He is negligible compared to
$^4$He($\alpha$n,$\gamma$)$^9$Be.

\end{abstract}


\section{Motivation}

Many calculations of the two--step radiative capture reactions of astrophysical
interest have been done by now in the framework of the single--particle
direct--capture model. An accuracy of the model was not known 
to a sufficient degree
and its assessment by performing a few--body type calculation
is of interest. This is accomplished in the present work for the 
$^4$He$(2n,\gamma)^6$He reaction.
  
In the past years a new neutron--rich astrophysical scenario taking place in the
neutrino--heated hot bubble between the nascent neutron star and
the overlying stellar mantle of a type--II supernova
has been proposed \cite{1}--\cite{5}.
The material,
originally in nuclear statistical equilibrium (NSE) at high temperature,
is expanded and cooled so rapidly that not all the $\alpha$--particles
have time to reassemble. In this environment most
nucleons are either in the form of free neutrons or bound in
$\alpha$--particles.
The nucleosynthesis in the neutrino bubble takes place in subsequent steps:
first, $\alpha$--particles are formed from the free nucleons. Then, in the
following $\alpha$--process, nuclei up to about $A \approx 100$
are produced \cite{1}. Finally, the neutrino bubble is also an ideal site
for the r--process
synthesizing the elements of about $ A \ge 100$ \cite{1,6,7}.

The bottleneck of the above nucleosynthesis processes is the formation of
nuclei with mass numbers $ A \ge 9$ from the nucleons
and $\alpha$--particles, because two--step processes have to be
involved in bridging the instability gaps at $A=5$ and $A=8$
(see Fig.~1).
Due to the lack of neutrons in hydrostatic
helium burning in red giants, these instability gaps have to be bridged
by the triple--alpha reaction $^4$He(2$\alpha$,$\gamma$)$^{12}$C
which proceeds as a two--step process $\alpha+\alpha \leftrightarrow ^{8}$Be
and $^8$Be($\alpha$,$\gamma$)$^{12}$C.
However, in the $\alpha$--process where neutrons are available the
instability gaps can also be overcome by the reaction
$^4$He($\alpha$n,$\gamma$)$^9$Be
which is much more efficient in the $\alpha$--process than
the triple--alpha reaction rate \cite{1,5}.
\begin{figure}
\vspace{30mm} 
\caption[Fig1]{Possible two--step processes
involved in bridging the instability gaps at $A=5$ and $A=8$. The main reaction
flow under typical conditions in the $\alpha$--process is indicated
by thick arrows, the thinner arrows represent almost negligible
reaction flows.}
\end{figure}

Another reaction chain
could also be of importance. This chain starts with
$^4$He(2n,$\gamma$)$^6$He 
and is then
continued by $^6$He($\alpha$,n)$^9$Be,
thereby closing the instability gaps at $A=5$ and $A=8$ (See. Here one
should take into account that the abundance of free neutrons for the 
$\alpha$--process is of the order of 10$^{20-30}$ neutrons cm$^{-3}$ 
\cite{1,6} and that the yield of the $\alpha 2n$ reaction forming $^6$He
is proportional to $Y_{\alpha}Y_n^2$ while the yield of the $2\alpha n$ 
reaction is proportional to $Y_{\alpha}^2Y_n$. (The notation $Y_{\alpha}$ 
and $Y_n$ stands for the concentrations of alphas and neutrons, respectively.) 
The rate of the $^4$He(2n,$\gamma$)$^6$He 
reaction was estimated in Ref. \cite{25}
and found to be very small. However, as it is shown below the main mechanism
of the reaction was missed in that study. We thus investigate whether 
the rate of the reaction is enhanced to a sufficient degree due to this main
mechanism under representative conditions. While we are using a three--body
model
of the process the same issue was studied in a recent
independent paper \cite{8} using 
a simpler direct--capture model.

The $^4$He(2n,$\gamma$)$^6$He--reaction is also of special interest 
because experimental data of the nucleus
$^6$He show features characteristic for halo nuclei \cite{9,10}.
This is indicated by
the special character of the $\alpha$--spectra for the decay of the
$J^\pi = 2^+$ level at $E=1.8$\,MeV in $^6$He and the
measurement of an abnormally large electromagnetic--dissociation
cross section \cite{11,12}. We note as well that the strength functions for
the reverse reaction were calculated in Ref. \cite{12a} in the framework of
an approach treating all the three final--state particles on equal footing.

\section{The Model}

Let us formulate our model for calculating the 
$^4$He(2n,$\gamma$)$^6$He reaction
rate. We adopt a two--step mechanism for the process.
In the first step the p--wave $3/2^{-}$ resonant state of $^5$He
is formed and the reaction proceeds through a wing of this resonance.
Calculations of
two--step reactions proceeding through the low--energy wing of a resonance
are well known from the triple--alpha process. However, the mechanism
of the reaction we are considering proves to be different.
In analogy with the triple--alpha process, a transition 
of a p--wave neutron to the resonant state of $^6$He
followed by a E2 capture to the ground state of $^6$He would be
the second step of the reaction. However there exists an
alternative possibility, namely a non--resonant E1--capture
of an s--wave neutron. The latter process proves to dominate strongly
the reaction in our case.
This occurs because of both the lower multipolarity and the absence of
a Coulomb or a
centrifugal barrier suppressing this transition. We thus adopt
this process as the second step of the reaction. The resonant
E2--contribution will be commented on below. We use the three--body
$\alpha$+n+n representation of the nuclear system in
our calculations.

The reaction rate per particle triplet for 
$^4$He(2n,$\gamma$)$^6$He, in
analogy to the triple--alpha process, is given by \cite{13,14}
\begin{equation}\label{e1}
\left<2{\rm n}^4{\rm He}\right> = 2 \int_{0}^{\infty}dE_1\,
\frac{\hbar}{\Gamma\left(^5{\rm He},E_1\right)}
\frac{d\left<{\rm n}^4{\rm He}\right>\left(E_1\right)}{dE_1}
\int_{0}^{\infty}dE_2\,
\frac{d\left<{\rm n}^5{\rm He}\right>\left(E_1,E_2\right)}{dE_2},
\end{equation}
where $E_1$ and $E_2$ are relative energies in the ($^4$He+n)--
and ($^5$He+n)--system, respectively. 
The quantity $\Gamma\left(^5{\rm He},E_1\right)$ is an
energy--dependent width of $^5$He,
whereas the integrands  
$d\left<{\rm n}^4{\rm He}\right>\left(E_1\right)/dE_1$
and
$d\left<{\rm n}^5{\rm He}\right>\left(E_1,E_2\right)/dE_2$ are 
transport cross sections
$\sigma(E)v$ \cite{15} for the
first and second step, 
respectively, times the corresponding Maxwell distributions 
$\sim\sqrt{E}\exp(-E/kT)$.

The first step of the reaction is 
represented by the first integral in Eq.~(\ref{e1})
\cite{13,14}.
An equilibrium 
between the production and the decay
for the population of $^5$He is assumed here.
The relative production rate of $^5$He 
is given by $\sigma_1(E_1)v_1$. The total production cross section 
$\sigma_1$ coincides with
the elastic ($^4$He+n) cross section 
as neutron emission of $^5$He is the only possible
exit channel. The relative decay rate
is given by $\Gamma\left(^5{\rm He},E_1\right)/\hbar$. The integrand is 
the ratio of the  $^5$He particle density at given $E_1$
to the product of net $\alpha$ and n particle densities 
at the equilibrium.
To calculate the
energy dependent width $\Gamma\left(^5{\rm He},E_1\right)$
we need a ($^4$He+n)--potential. It was obtained by scaling the
depth and range of the folding potential to the experimental
energy and width of the  $^5$He ground state. Both these quantities
are exactly reproduced by our potential and $\sigma_1(E_1)$ is also reproduced 
reasonably well.    The values 0.89\,MeV
\cite{16} and 0.76\,MeV were adopted as the
Breit--Wigner $^5$He energy and width, respectively.  

The value of the width 
was deduced \cite{17} from the most recent experiment \cite{18}.
In Ref.~\cite{18} itself a significantly larger value of the width was
reported and we would like to explain the difference between the two
values. In Ref.~\cite{18} the spectrum of that work was fitted by the 
expression of the form
\begin{equation}\label{width}
\frac{\Gamma_R(E_{{\rm n}\alpha})}{[E_{{\rm n}\alpha}-E_{\rm res}-\Delta
(E_{{\rm n}\alpha})]^2+\left[\frac{\Gamma_R(E_{{\rm n}\alpha})}{2}\right]^2}
\end{equation}
times a smooth function in the $({\rm n}-\alpha)$ relative energy
$E_{{\rm n}\alpha}$. Here $\Gamma_R(E)=2\gamma^2P(E)$ and 
$\Delta(E)=-\gamma^2[S(E)-S(E_{\rm res})]$ where $\gamma^2$ is the
reduced width, and $P$ and $S$ are the penetrability and the shift
function defined as usual. The value of $\Gamma_R(E_{\rm res})$ was then
reported as a width. However, this $R$--matrix value is known to differ
from the physical width determining the lifetime of a system.  Just the 
latter width $\Gamma$ determines the energy-dependent width in Eq. (\ref{e1}) 
and it can be obtained from
the location of the pole $E_{{\rm n}\alpha}=E_0-i\Gamma/2$ 
of the factor (\ref{width}). This width
was found in 
Ref.\cite{17} and the value obtained is close to the older
estimates \cite{16}.

To calculate the second step reaction rate $\left<{\rm n}^5{\rm
He}\right>$, the $^5$He(n,$\gamma$)$^6$He cross section is
required.  As explained above, we consider the electric dipole
transition E1 from an incident s--wave. The cross section is given by
\begin{equation}\label{e2}
\sigma_2\left(E_1,E_2\right) = \frac{2\pi}{81} \left( \frac{E_\gamma}{\hbar c}
\right)^3 \frac{e^2}{\hbar v_2} \mid I \mid ^2
\end{equation}
with
\begin{equation}\label{e3}
I = \frac{1}{k_2} \int_0^\infty \!\! dr_1 \int_0^\infty \!\! dr_2
u_{\rm b}(r_1) r_1 u_{\rm sc} (E_2,r_2) r_2^2  f_{\ell=1,j=3/2,} 
(r_1,r_2) \quad ,
\end{equation}
where $E_\gamma=E_1+E_2+Q_{12}$ is the energy of the photon, with 
$Q_{12}=0.98$\,MeV being the
Q--value of the reaction $^4$He(2n,$\gamma$)$^6$He. In Eq.~(\ref{e3})
the coordinates
$r_1$ and $r_2$ are the distances between the valence neutron  
and the $\alpha$--core
or the
center of mass of $^5$He, respectively.
The quantities $v_2$ and $k_2$ are the relative velocity and wave number 
in the entrance channel
($^5$He+n) of the second step.
The quantity
$u_{\rm b}(r_1)/r_1$ represents the radial part of the
quasi--bound wave function of the $^5$He resonance, and
$u_{\rm sc} (E_2,r_2)/r_2$ is the radial part of
the ($^5$He+n) scattering wave function. The function
$f_{\ell=1,j=3/2} (r_1,r_2)$ is the radial component in the
expansion
\begin{equation}\label{e4}
\Psi\left(^6{\rm He}\right) = \sum_{j\ell} f_{\ell j}(r_1,r_2)
\left[\chi_{\ell j}\left({\hat{\bf r}}_1,\sigma_{1z}\right)
\chi_{\ell j}\left({\hat{\bf r}}_2,\sigma_{2z}\right)\right]_{J=0}
\end{equation}
of the $\alpha$+n+n ground--state wave function.
Here $\chi_{\ell j}$ are angular--mo\-men\-tum functions
with orbital momenta $\ell$ (necessarily equal to each other) and total
angular momenta $j=\ell \pm 1/2$,  $\sigma_{1z}$ and  $\sigma_{2z}$
are spin variables of the valence neutrons, and the brackets $[\ldots]$
denote vector coupling.  Only the
$(\ell=1,j=3/2)$--component may contribute to the cross section. The
$^6$He wave function is
normalized to unity and the scattering wave function  
is normalized as $u_{\rm sc} \rightarrow \sin(kr_2+\delta)$ in
the asymptotic region.

The ground--state wave function $u_{\rm b}(r_1)$ of
$^5$He
was obtained as the Siegert solution to the ($^4$He+n) Schr\"odinger
equation and this solution was taken inside the centrifugal barrier and 
normalized there to unity (see e.g.~\cite{19}). The Siegert solution is 
an eigenfunction corresponding to the
boundary condition for a decaying system and inside the barrier it represents
the contribution of a resonance to continuum wave functions.
Its absolute value 
decreases inside the barrier and increases in the
outer region.  
The
cut--off radius 7.425\,fm was taken at the point of a minimum of $\mid
u_{\rm b}\mid$. Replacement of this
complex--energy ($E_0-i\Gamma/2$) solution by the real--energy
$E_0$ scattering solution with the same cut--off radius leads to
less than only 1\% difference in the second--step cross section.
This shows both that the 
resonance
contribution is a predominant one and that the cut--off radius has
been 
chosen correctly.  Variation of
this radius, e.g., by $\pm 1.5$\,fm changes the cross section by not more than 
20\%. (Reasonable variations of the cut--off radii give even
narrower limits.) The above-mentioned
($^4$He+n) folding potential was used to 
obtain $u_{\rm b}$.
  
To calculate the scattering wave function $u_{\rm sc} (E_2,r_2)$
we construct the optical ($^5$He+n)--potential
by using the folding procedure \cite{20,21}.   We
first reproduced the matter density distribution for $^4$He by
means of a simple Gaussian with the r.m.s.~matter radius of 1.469\,fm
deduced from the r.m.s.~charge radius of 1.676\,fm \cite{22}.  To
obtain the density distribution for $^5$He, we folded the
$^4$He--density  distribution with the above--mentioned wave function
$u_{\rm b}(E_1,r_1)$.  Finally, we used again the folding procedure
to obtain the potential. Due to abnormally large separations between
the valence neutrons and the $\alpha$ core in the  final--state $^6$He
halo nucleus the cross section is largely
enhanced. The outer free-motion parts of the $u_{\rm b}$ and 
$u_{\rm sc}$ functions contribute
predominantly to Eq. (\ref{e3}) due to the same reason.
If one replaces the ($^4$He+n) potential we used above by
the square--well potential whose range and depth are fitted to the
position and width of the  $^5$He resonance the reaction rate
obtained changes by
less than 1\% and thus it is insensitive to the
form of the potential.
An accurate three--body wave function of
$^6$He (see \cite{10}) was used
in our calculations. Originally, it
was given in $LS$--coupling and in the form
of an expansion over hyperspherical harmonics and we present it
in the form of Eq.~(\ref{e4}).
 
While we included the formation of the $^5$He$(3/2^-)$ resonance as
the first step of the reaction we neglected the contribution from the
broad $1/2^-$ resonance. Taking into account Eq.~(\ref{e1}) the relative
$1/2^-$ contribution may be roughly estimated as
\begin{equation}\label{1/2est}
[\sigma_{1/2}(E)\Gamma_{3/2}w_{1/2}][\sigma_{3/2}(E)\Gamma_{1/2}
w_{3/2}]^{-1}
\end{equation}
where $\sigma$'s are elastic cross sections, $\Gamma$'s are widths and
$w$'s are weights
of the $f_{\ell=1, j}$ components with $j=1/2$ and 3/2 in the
$^6$He wave function from Eq.~(\ref{e4}). The weights we
calculated are $w_{3/2}=0.87$ and $w_{1/2}=0.049$. The component
entering Eq.~(\ref{e3}) thus dominates the $^6$He wave function.
Substituting
these values into (\ref{1/2est}) along with the value of
$\Gamma_{1/2}\simeq 4$ MeV \cite{16} we obtain
$0.01 \sigma_{1/2}(E)/\sigma_{3/2}(E)$.  The ratio of the cross sections
is less than unity. Thus the
contribution of the  $^5$He$(1/2^-)$  resonance to the reaction
rate is negligible.

\begin{figure}
\vspace{30mm} 
\caption[Fig2]{Calculated cross section for $^5$He(n,$\gamma$)$^6$He}
\end{figure}
\begin{figure}
\vspace{30mm} 
\caption[Fig3]{Calculated cross sections for $^6$He($\gamma$,n)$^5$He}
\end{figure}

\section{Reaction Rates}

The calculated cross section for $^5$He(n,$\gamma$)$^6$He
can be parametrized by
$\sigma = 0.152/\left(E_2\,[{\rm
{MeV}}]\right)^{1/2}$\,[$\mu$b] and it is depicted in Fig.~2. Furthermore, in
Fig.~3 the cross section for the inverse process
$^6$He($\gamma$,n) $^5$He, as obtained from the principle of
time--reversal invariance \cite{23,24}, is shown. The cross
sections given in these figures were obtained by assuming that
$^5$He exists exactly
at the resonance energy $E_1 = 0.89$\,MeV.
The $^5$He(n,$\gamma$)$^6$He cross section exhibits the
$1/v$--behavior corresponding to incident s--waves.
The $^6$He($\gamma$,n)$^5$He cross section
shown in Fig.~3 is about three
orders of magnitude higher than the maximum value that was originally used 
for the determination of the
reaction rates of $^4$He(2n,$\gamma$)$^6$He
and $^6$He($\gamma$,2n)$^4$He
by Fowler et al.~\cite{25} (cf. below).
\begin{figure}
\vspace{30mm} 
\caption[Fig4]{Ratio of calculated reaction rate of 
$^4$He(2n,$\gamma$)$^6$He
for $T_9= 0.5-8.0$ and the maximum of the previous adopted value~\cite{25}}
\end{figure}
\begin{table}
\caption{Parameters of the reaction rate for $^4$He(2n,$\gamma$)$^6$He
calculated in our three--body model}
\begin{tabular}{rrr}
\hline
& $0.1 \le T_9 \le 2$ & $2 < T_9 \le 15$\\
\hline
a & 0.00265 & 0.293 \\
b & 2.55 & $-$0.351\\
c & 0.181 & $-$5.24 \\
\hline
\end{tabular}
\end{table}

The obtained reaction rate can be parametrized in
the following way:
\begin{equation}\label{e5}
N_{\rm A}^2 \left<2{\rm n}^4{\rm He}\right> = a
(T_9)^b\exp\left(\frac{c}{T_9}\right)\,
10^{-8}\,{\rm{cm}}^6\,{\rm s}^{-1}\,{\rm {mol}}^{-2}
\quad ,
\end{equation}
where the parameters $a,b,c$ are listed
in Table 1 for the two temperature regions
$0.1 \le T_9 < 2$ and $2 \le T_9 \le 15$ ($T_9$: in units of
$10^9$\,K). 

The inverse reaction rate $\lambda_\gamma$ per nucleus per second
can be calculated by using the RevRatio \cite{25}
\begin{equation}\label{e5a}
\lambda_\gamma={\rm RevRatio}\times N_{\rm A}^2 
\left<2{\rm n}^4{\rm He}\right>\,
{\rm s^{-1}}  
\quad .
\end{equation}
The RevRatio is parametrized in the following way
\begin{equation}\label{e5b}
{\rm RevRatio}=1.08 (T_9)^3 \exp\left(\frac{-11.3}{T_9}\right)\, 
10^{20}\,{\rm cm^{-6}\,mol^{2}}
\quad .
\end{equation}

As shown in Fig.~4, the reaction rate for
$^4$He(2n,$\gamma$)$^6$He calculated with the
help of our three--body
model is more than three orders of magnitude larger than the
maximum of the previously adopted value~\cite{25}. This is mainly due to the
non--resonant E1--transition of the second step 
$^5$He(n,$\gamma$)$^6$He,
which dominates the cross section, and which was not taken
into account in Ref.~\cite{25}. 
We have also obtained an estimate
comparable to the results in Fig.~4 
using the direct--capture model.

The huge enhancement of the reaction rate obtained can 
simply be demonstrated as follows. Let us estimate the ratio of the 
second--step non--resonant 
transport cross section 
$\langle\sigma v\rangle_{nr}$ to the resonant one. To simplify the reasoning, 
let us consider 
the case of higher temperatures such that $kT$
is comparable to or larger than the width $\Gamma_1$ of $^5$He
($T$ larger than about $10^{10}$\,K).
Below it will be convenient to treat the 
$\langle\sigma v\rangle_{nr}$ value as an average
transition 
probability. This implies that the $^5$He+n 
relative wave function is normalized to $\exp(i{\bf kr})$ plus the 
scattered wave at large distances. 
The resonant second--step transition 
disregarded in our calculations proceeds through the 
first excited
state $^6$He$(2^+)$  lying within the width of the broader
$^5$He resonance. The  quantity $\langle\sigma v\rangle$ for 
this transition can be
estimated as (c.f.~\cite{15})
$(\hbar^2/2\mu)\mu^{-1/2}\Delta^{-3/2}x^{3/2}\exp(-x)\hbar w_r$ 
where $\Delta$ is the position of the $^6$He resonance
with respect to the ($^5$He+n) threshold, $x=\Delta/kT$,
$\mu$ is the reduced mass for ($^5$He+n)
and $w_r$ is the transition probability pertaining to the
transition from the $^6$He$(2^+)$ quasi--bound state to the ground state. The 
quantity
$\Delta$ equals to $(E_r-E_1)$ where $E_r$ is the absolute position of the
$^6$He$(2^+)$ resonance and $E_1$ is the energy of  $^5$He. This
quantity should be averaged over the distribution of $E_1$ giving 
$\bar{\Delta}(T)\simeq \Gamma_1$.
The ratio $\langle\sigma v\rangle_{nr}/w_r$ of the two transition 
probabilities equals to
the squared ratio of absolute values of electromagnetic transition matrix 
elements. The transitions
considered are E1 and
E2, respectively, and this ratio can thus be estimated
as $R^{3}[(2l+1)/k_{\gamma}R]^2$ where $l=2$, 
$k_{\gamma}\simeq 10^{-2}$\,fm$^{-1}$ is the photon momentum, 
$R\simeq 4.5$\,fm
\cite{10} is the average distance
between a ($^4$He+n) subsystem and the other outer neutron in the 
$^6$He nucleus and the
$R^{3}$ factor comes from the difference in normalizations
of the initial state wave functions entering the transition matrix elements.
As a result we arrive at the rough estimate 
\[25\left[\frac{(\hbar k_{\gamma})^2}{2\mu}\right]^{-1}
\left(\frac{\hbar^2}{\mu R^2}\right)^{-1/2}\Gamma_1^{3/2}
\bar{x}^{-3/2}\exp(\bar{x})\simeq 5\cdot 10^{3}\bar{x}^{-3/2}\exp(\bar{x})\]
for the ratio of the reaction rates, where $\bar{x}\simeq\Gamma_1/kT$.

The rate obtained in our three-body calculation 
is about a factor of 1.6 smaller
than the simpler single--particle direct--capture rate given in 
Ref.~\cite{8}.
Therefore, we reach the same conclusion as in Ref.~\cite{8}, i.e.
for the trajectories relevant to
the r--process given in Ref.~\cite{4} (the last
16 trajectories given in Table III) the 
reaction rate of the $^4$He(2n,$\gamma$)$^6$He obtained
is stronger by a factor of three than the triple--$\alpha$
reaction, but is still weaker by approximately two orders of
magnitude than the $^4$He($\alpha$n,$\gamma$)$^9$Be
reaction.

With respect to the photodisintegration of $^6$He
our calculations show that for the same trajectories as given above, 
the beta decay of $^6$He
and $^6$He(2n,$\gamma$)$^8$He are 
slower by about 10 and 11 orders of
magnitude, respectively
(the reaction rate for $^6$He(2n,$\gamma$)$^8$He
is given in \cite{26}).
The reaction $^6$He($\alpha$,$\gamma$)$^{10}$Be
is smaller by about two orders of magnitude.
The only reaction which can compete with the photodisintegration is
the reaction $^6$He($\alpha$,n)$^9$Be.
The cross section for the inverse reaction 
$^9$Be(n,$\alpha$)$^6$He was measured by Stelson and
Campbell \cite{27} in the energy range from 0.7 to 4.4 MeV.
From this the reaction rate of
$^6$He($\alpha$,n)$^9$Be was deduced. 
A comparison of the rate with the photodisintegration rate is
given in Table 2 for different temperatures. 
While the rate for the photodisintegration
is multiplied with the density $\rho$ in units g\,cm$^{-3}$ and the
$^6$He abundance in the
network, the rate for the $\alpha$ capture is multiplied with the
square of the density and the abundances of $^6$He and
$\alpha$. Therefore, under representative conditions (cf., Ref.~\cite{1})
with $\rho = 5 \times 10^4$\,g\,cm$^{-3}$ and $Y_{\alpha} = 0.225$
most of the produced $^6$He
can be processed further on to $^9$Be. Even at $T_9 = 5.0$
alpha capture by $^6$He is still more important than photodisintegration.
However, it has to be noted that under representative conditions
in the $\alpha$--process the reaction path proceeding
through $^6$He is still
negligible compared to the reaction
$^4$He($\alpha$n,$\gamma$)$^9$Be.

\begin{table}
\caption{Comparison of the reaction rates of the photodisintegration
$^6$He($\gamma$,2n)$^4$He (second column) and 
$^6$He($\alpha$,n)$^9$Be (third column) for different
temperatures (first column)}
\begin{tabular}{ccc}
\hline
$T_9$ & $\lambda_\gamma$  & $N_A <^{6}$He$\alpha>$ \\
\hline
0.5 & $1.28 \times 10^{-2}$ & $6.45 \times 10^5$ \\
0.8 & $7.71 \times 10^2$ & $3.16 \times 10^6$ \\
1.0 & $4.36 \times 10^4$ & $6.69 \times 10^6$ \\
1.5 & $1.66 \times 10^7$ & $2.64 \times 10^7$ \\
2.0 & $4.87 \times 10^8$ & $6.86 \times 10^7$ \\
2.5 & $4.54 \times 10^9$ & $1.37 \times 10^8$ \\
3.0 & $2.27 \times 10^{10}$ & $2.26 \times 10^8$ \\
5.0 & $8.50 \times 10^{11}$ & $6.05 \times 10^8$ \\
\hline
\end{tabular}
\end{table}

In conclusion, we have considered a new mechanism of the two--neutron
capture by the $^4$He nucleus which leads to a more than three order of
magnitude enhancement with respect to the value adopted by Fowler et al.
\cite{25}. 
We have performed a simplified three--body calculation of the reaction rate
that proved to be in accordance with the prediction of the simple
direct--capture
model within a factor of two. The inverse reaction of the photodisintegration
of $^6$He has also been studied. Our cross--section for the photodisintegration
via formation of the intermediate $^5$He state can be verified in experiments 
on $^6$He break--up on heavy targets at low energies, of the type performed 
last years \cite{50,51} for the $^{11}$Li nucleus.  
The results obtained have a relevance to other 2N radiative capture 
processes as well.

The authors are indebted to I.J.~Thompson
for help in their work and to J.S. Vaagen and M.V. Zhukov for valuable
discussions on halo nuclei.
Useful discussions with J.~G\"orres, M.~Wiescher and M.J.~Balbes
are acknowledged. This work was supported by the
Fonds zur F\"orderung der wissenschaftlichen
For\-schung in \"Osterreich (project P10361--PHY), by the
\"Oster\-rei\-chi\-sche Nationalbank (project 5054) and by the International
Science Foundation and Russian Government (grant J4M100).

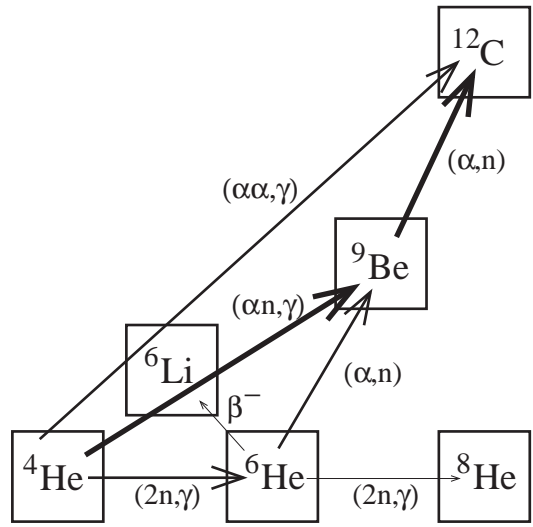
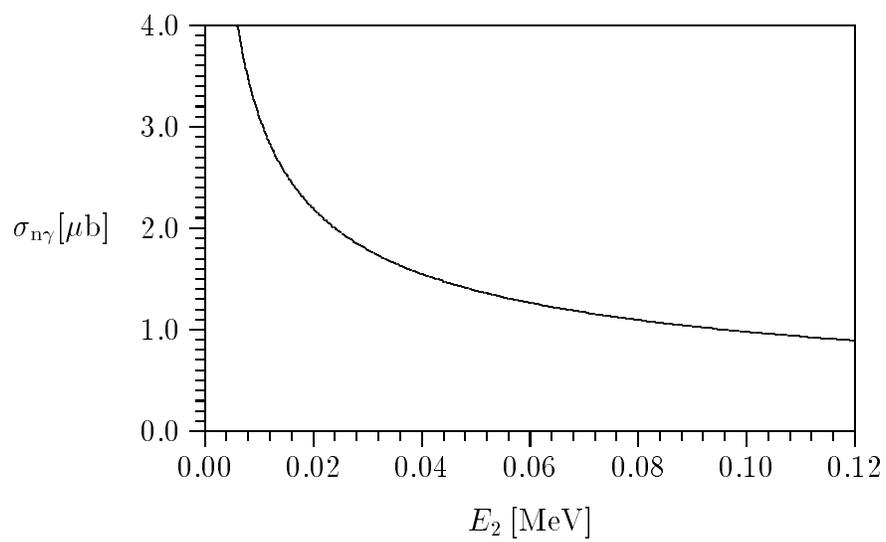
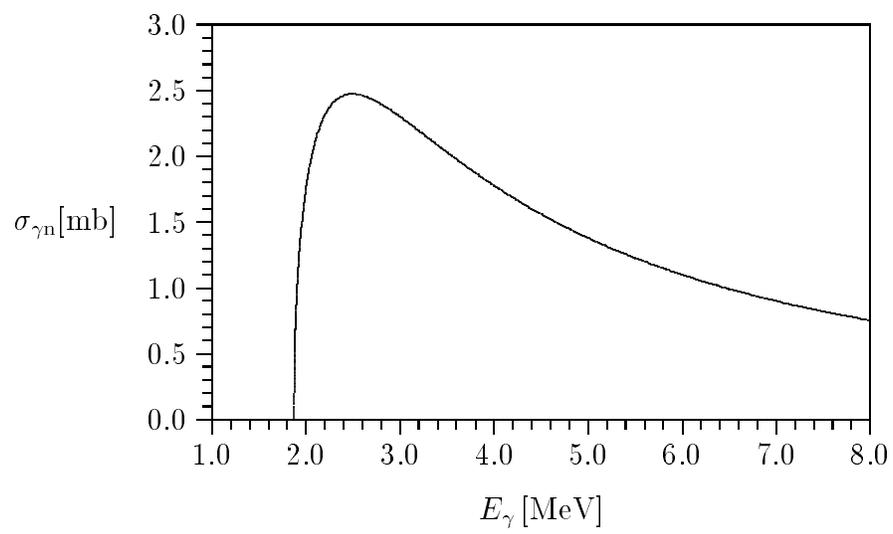
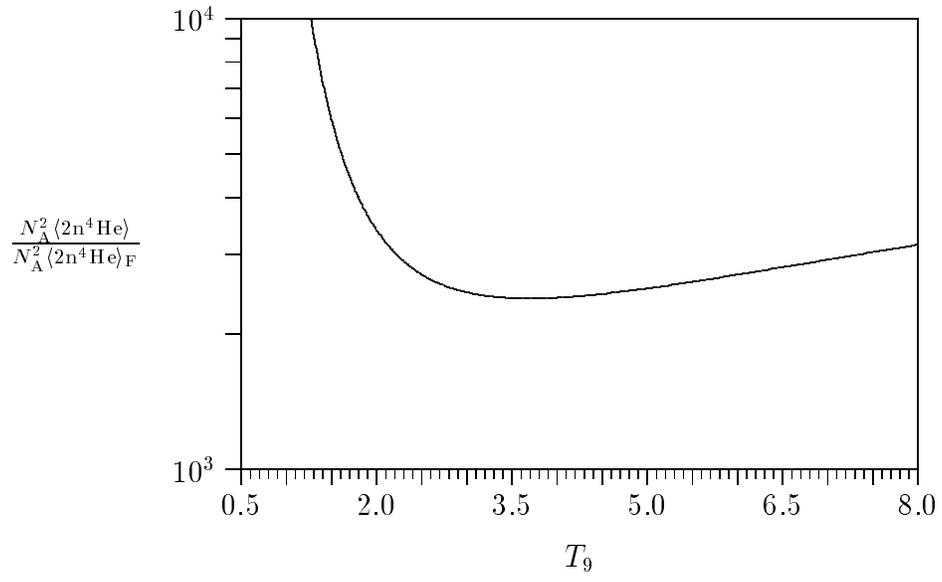
\end{document}